\documentclass[twocolumn,aps]{revtex4-2}
\usepackage{amsmath}
\usepackage{amssymb}
\usepackage{graphicx}
\usepackage{dcolumn}
\usepackage{bm}
\usepackage{xcolor}
\usepackage{epstopdf}
\usepackage{longtable}

\begin{document}

\preprint{Phys.Rev.B }

\title{Electron Hydrodynamics and Bernoulli Effect in Venturi-Shaped 2D Systems}
\author{C. A. Monari$^1$ , A. D. Levin,$^1$   A. S. Jaroshevich,$^{2}$  Z. D. Kvon,$^{2,3}$  V. A. Chitta,$^1$ D. V. Dmitriev,$^{2}$ A. K. Bakarov$^{2}$ and G. M. Gusev$^1$}

\affiliation{$^1$Instituto de F\'{\i}sica da Universidade de S\~ao
Paulo, 135960-170, S\~ao Paulo, SP, Brazil}
\affiliation{$^2$Institute of Semiconductor Physics, Novosibirsk
630090, Russia}
\affiliation{$^3$Novosibirsk State University, Novosibirsk 630090,
Russia}

\date{\today}
\begin{abstract}
The study of electron hydrodynamics provides a powerful framework for understanding transport in ultraclean conductors, yet experimental evidence has thus far been largely restricted to the linear response regime. Here, we report the direct observation of a strongly nonlinear transport regime in a high-mobility two-dimensional electron system. By engineering devices in a Venturi-shaped (wedge) geometry specifically designed to enhance convective nonlinearities, we uncover a pronounced nonlinear voltage response and a large diodicity in the current–voltage characteristics. Our experimental findings show quantitative agreement with a theoretical model that attributes the observed nonlinearity to the convective acceleration of the electron fluid, analogous to the Bernoulli effect. These results provide compelling evidence for the applicability of the hydrodynamic framework to electron transport in two dimensions and open new avenues for exploring nonlinear and preturbulent phenomena in solid-state systems.
\end{abstract}

\maketitle

\section{Introduction}
The hydrodynamic description of electrons in two-dimensional fermionic systems provides a powerful alternative to conventional kinetic theory, offering new insights into transport behavior at the microscale. When electron–electron interactions dominate, the collective motion of carriers can be effectively captured within a viscous fluid framework, giving rise to unconventional transport signatures. Advances in material fabrication—particularly the ability to produce ultraclean samples—have recently made it possible to probe these hydrodynamic effects in a controlled manner across diverse two-dimensional platforms.

Hydrodynamic flow regimes are expected to emerge when the mean free path for electron–electron scattering ($l_{ee}$) becomes much shorter than the mean free path associated with momentum-relaxing processes such as impurity and phonon scattering ($l$). Over the past decade, this perspective has sparked growing interest in solid-state physics, leading to a wealth of theoretical predictions and their subsequent experimental verification \cite{narozhny, hui}.

Several hydrodynamic analogs have been predicted and observed in electronic fluids based on two-dimensional systems such as graphene and GaAs quantum wells. The first is the Poiseuille-like velocity profile in narrow channels, which is a hallmark of viscous liquid flow. Due to the parabolic profile, the maximum velocity increases with viscosity, leading to a decrease in conductivity as the temperature rises—known as the Gurzhi effect \cite{gurzhi, dejong, andreev, narozhny2, gusev1, gusev2, principi, keser}. This Poiseuille-like profile has been directly visualized in graphene samples \cite{sulpizio} and indirectly detected through measurements of recombination times in GaAs narrow channels \cite{patricio}. Another hydrodynamic signature is the formation of current whirlpools, which has been indirectly confirmed by negative resistance measurements arising from current backflow \cite{bandurin, levin, braem}, and directly imaged in graphene using nanoscale scanning magnetometry \cite{palm}. A further analogy is the Stokes flow around  obstacles, investigated both through transport measurements \cite{lucas, gusev3, levin2} and via scanning tunneling potentiometry, which enabled direct visualization and mapping of the spatial variation of the electrochemical potential across a channel formed between two electrostatic barriers \cite{krebs}. 

The decrease in viscosity with increasing magnetic field, as observed in classical plasmas \cite{braginskii}, leads to giant negative magnetoresistance—predicted in \cite{alekseev} and experimentally observed in narrow GaAs mesoscopic channels \cite{gusev1, gusev4, levin3, wang, wang2, wang3}.

Studies of electron hydrodynamics have so far concentrated primarily on the linear Stokes--Ohm regime, where transport phenomena remain relatively simple. A central objective, however, is to reach the nonlinear domain, where richer fluid-like effects can emerge. In classical fluids, nonlinearities typically become significant at high flow velocities, quantified by large Reynolds numbers. For electronic systems, this presents a major obstacle, since state-of-the-art devices generally operate at extremely low Reynolds numbers ($\mathrm{Re}\sim 10^{-2}$), making nonlinear behavior difficult to access under standard conditions. In this context, much of the previous discussion of nonlinear hydrodynamic effects has been associated with higher-Reynolds-number or pre-turbulent regimes, for example in channel flows where the onset of turbulence is expected to require substantially larger Reynolds numbers, wider channels, or additional constrictions and obstacles~\cite{Erdmenger}.

One promising pathway to bypass this limitation is the electronic analogue of the Bernoulli effect \cite{hui,hui2}. This effect originates from the convective acceleration of the electron fluid and is highly sensitive to geometry. In particular, it is designed to manifest in channels with a varying cross-section—such as those with a Venturi (wedge-shaped) profile—while it vanishes entirely in conventional rectangular geometries where the width is constant.

The outcome of this mechanism is a uniquely nonlinear current–voltage ($I$–$V$) response predicted in the paper \cite{hui2}. A defining feature of the nonlinearity is its even symmetry with respect to current reversal, which makes it experimentally distinguishable from other transport mechanisms. This symmetry allows for a direct probe through the symmetrized current, defined as $I_{\mathrm{sym}}(V_{0}) \equiv \tfrac{1}{2}\,\big[ I(V_{0}) + I(-V_{0}) \big]$. A non-zero $I_{\text{sym}}$ provides a clear and unambiguous signature of the Bernoulli-like effect. Observing such a strong, geometry-driven nonlinear response would therefore constitute a distinct hallmark of the hydrodynamic regime, setting it apart from conventional diffusive transport, where no such effect can occur.

In this work, we demonstrate the direct realization of a strongly nonlinear transport regime in high-mobility two-dimensional electron systems hosted in narrow and wide GaAs quantum wells. To access this regime, we designed devices with a Venturi-shaped (wedge) geometry specifically intended to enhance convective nonlinear effects. This geometry gives rise to a strong nonlinear voltage response together with a pronounced diode-like asymmetry in the current--voltage characteristics. The experimental results are in good quantitative agreement with a theoretical description that attributes the nonlinearity to convective acceleration of the electron fluid, providing an electronic analogue of the Bernoulli effect.

A phenomenologically related effect was reported earlier in a three-terminal GaAs electron jet pump, where negative side current and current amplification were described as an ``electronic Venturi effect''~\cite{Taubert}. However, that work interpreted the effect in terms of nonequilibrium hot-electron scattering and charge separation, rather than within a hydrodynamic Bernoulli framework.
\section{EXPERIMENTAL RESULTS}
\begin{figure}[ht]
\includegraphics[width=8cm]{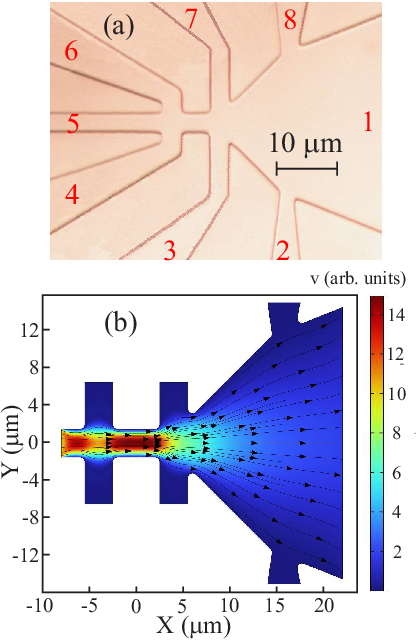}
\caption{\label{sample}(Color online)
 (a) Optical image of the device containing both a narrow rectangular channel and a Venturi-shaped channel. (b) Calculated hydrodynamic flow pattern, showing the velocity distribution in the device.}
\end{figure}
The single-layer devices were fabricated from high-quality GaAs quantum wells of nominal width 16~nm, with an electron density of approximately $9.0 \times 10^{11}~\mathrm{cm}^{-2}$ at 4.2~K. The corresponding macroscopic wafer exhibited an electron mobility of about $2 \times 10^{6}~\mathrm{cm}^2/\mathrm{V\,s}$. The bilayer devices were fabricated from wide GaAs quantum wells, in which charge redistribution across the well forms a bilayer electron system with two occupied subbands separated by a soft electrostatic barrier. For such bilayer structures, a representative wide-well sample has a width of 46~nm, a total electron density of $6.7 \times 10^{11}~\mathrm{cm}^{-2}$, and a mobility of about $2 \times 10^{6}~\mathrm{cm}^2/\mathrm{V\,s}$ at 4.2~K. In comparison with a single-layer system, the bilayer configuration allows additional intersubband scattering and generally stronger electron--electron scattering, which can promote hydrodynamic transport and modify the boundary conditions through a shorter slip length \cite{gusev2}

Transport measurements were performed on a multiterminal mesoscopic device with a Venturi-shaped channel. As seen in Fig.~1(a), the central part of the structure consists of a narrow constriction that expands into a wedge-like section, forming an electronic analogue of a Venturi nozzle. Such a geometry is expected to enhance current focusing and to strengthen hydrodynamic convective effects.
For comparison with theory, the Venturi section is parameterized by an opening angle $\theta_0$ and two radial coordinates, $r_0$ and $r_1$, measured from the apex of the wedge to the entrance and exit cross sections of the active channel, respectively. The corresponding channel widths are $h_0=\theta_0 r_0$ and $h_1=\theta_0 r_1$, while the length of the active Venturi region is $l_r=r_1-r_0$.

The geometric parameters of the fabricated device were estimated from the optical image. As a first approximation, the active Venturi region can be represented as a circular wedge. In this construction, the entrance and exit of the wedge-shaped section are located at approximately \(r_0\approx5~\mu\mathrm{m}\) and \(r_1\approx20~\mu\mathrm{m}\) from the apex, giving an effective Venturi length \(l_r\approx15~\mu\mathrm{m}\). Taking the opening angle to be \(\theta_0\approx\pi/2\), one obtains the effective wedge widths \(h_0=\theta_0 r_0\approx7.85~\mu\mathrm{m}\) and \(h_1=\theta_0 r_1\approx31.4~\mu\mathrm{m}\), with \(h_1/h_0\approx4\). We emphasize, however, that these values should be regarded only as effective geometrical estimates of an idealized wedge, rather than an exact representation of the fabricated multiterminal device.

For comparison with the actual lithographic structure, the physical channel width changes gradually from a narrow entrance width \(W_{\rm in}\simeq3~\mu\mathrm{m}\) to a wider exit width \(W_{\rm out}\simeq22~\mu\mathrm{m}\) over approximately the same active Venturi length, \(l_r\simeq15~\mu\mathrm{m}\). In the quantitative estimates below, we therefore treat \(W_{\rm in}\), \(W_{\rm out}\), and \(l_r\) as effective parameters of the active Venturi region, while keeping in mind that finite slip, current spreading, edge depletion, and the nonideal lithographic shape may renormalize the effective hydrodynamic width of the constriction.

 Ohmic contacts to the two-dimensional electron system were created
by annealing Ti/Ni/Au layers deposited on the GaAs surface.  Three single-layer and two bilayer devices, all fabricated from the same substrates, were studied under the same experimental conditions.  All devices demonstrated consistent and reproducible results.

\begin{figure}[ht]
\includegraphics[width=8cm]{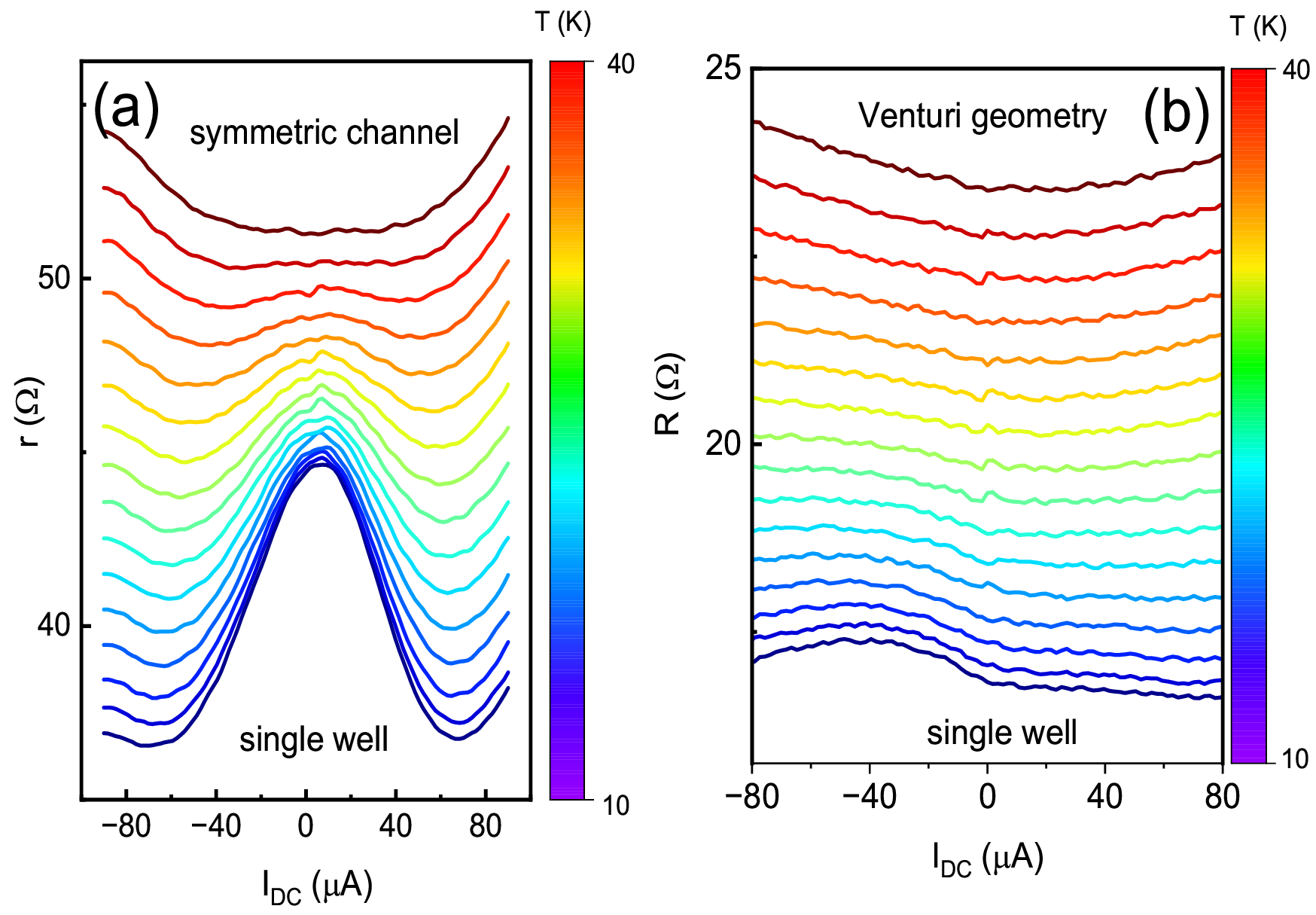}
\caption{\label{sample}(Color online)
(a) Evolution of the differential resistance as a function of applied DC current and temperature for the symmetric long-channel geometry, Single layer. (b) Evolution of the differential resistance as a function of applied DC current and temperature for the Venturi-geometry channel, Single layer }
\end{figure}

\begin{figure}[ht]
\includegraphics[width=8cm]{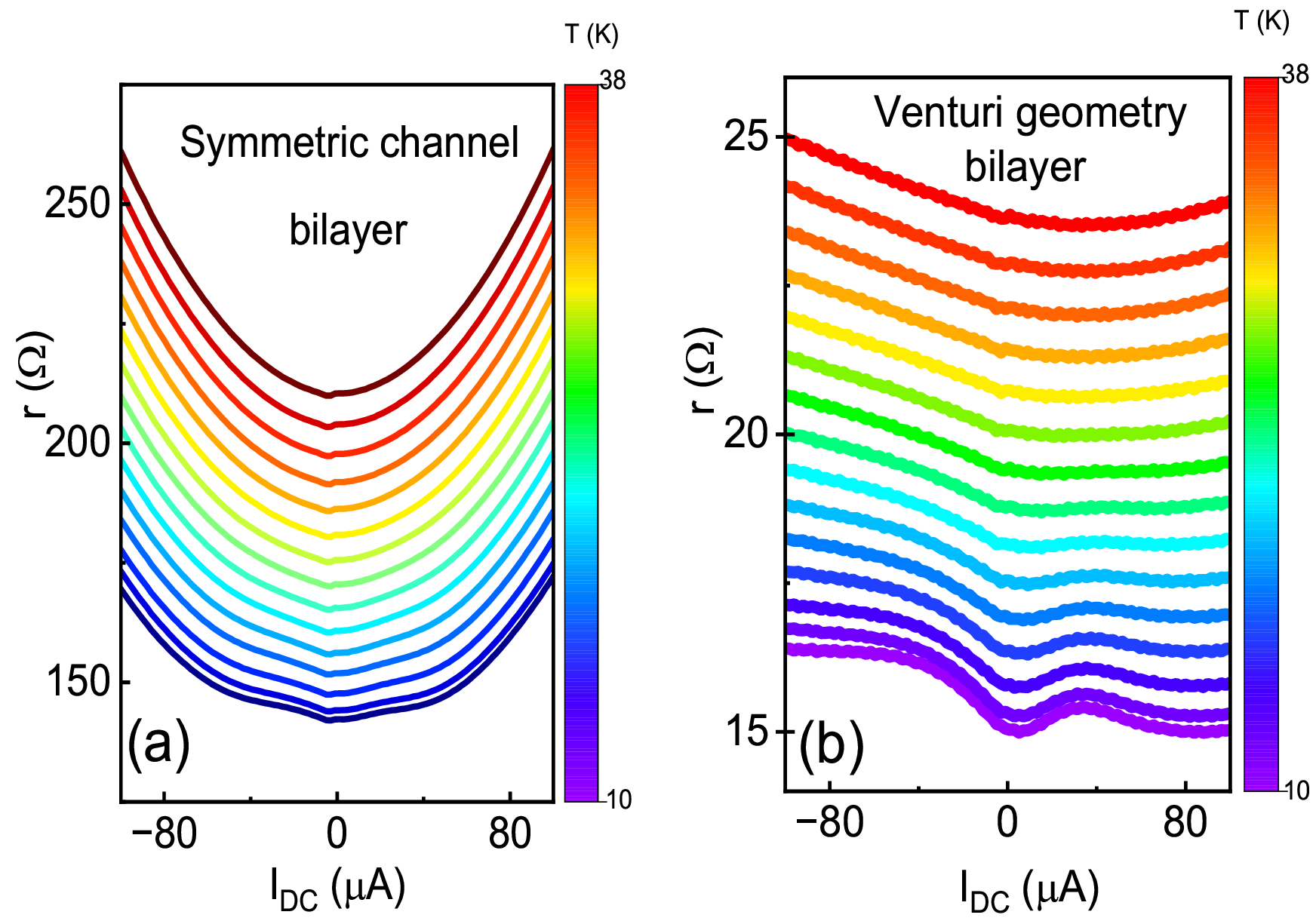}
\caption{\label{sample}(Color online)
(a) Evolution of the differential resistance as a function of applied DC current and temperature for the symmetric long-channel geometry, bilayer. (b) Evolution of the differential resistance as a function of applied DC current and temperature for the Venturi-geometry channel, bilayer}
\end{figure}

The measurements were carried out in a variable-temperature-insert cryostat using a standard lock-in technique to record the longitudinal resistance. To minimize electron overheating, a low-frequency alternating excitation current in the range from 0.1~$\mu$A to 1~$\mu$A was applied. In addition, a direct current, $I_{\mathrm{dc}}$, was superimposed through the same current leads in order to determine the differential resistance, defined as $r = r_{xx} = \frac{dV_{xx}}{dI_{\mathrm{dc}}}$. The experiments were performed in a multiterminal configuration adapted to the Venturi-shaped structure in order to emphasize hydrodynamic transport effects~\cite{hui2}. In this arrangement, the current was driven through the main channel of the device, while the voltage drop was measured between the appropriate side probes, yielding the four-terminal resistance. The current was applied between contacts 1 and 5, while the voltage $V$ was measured across probes 7 and 6, yielding the zero-bias resistance
$R = R_{1,5}^{7,6} = \frac{V_{7,6}}{I_{1,5}}$ for the symmetric channel [see Fig.~1(a)]. In addition, we performed measurements in a Venturi configuration. In this case, the current $I$ was applied between contacts 1 and 5, while the voltage $V$ was again measured across probes 8 and 7, yielding $R = R_{1,5}^{8,7} = \frac{V_{8,7}}{I_{1,5}}$.

The main objective of this work is to investigate the zero-field resistivity as functions of temperature and applied direct current, $I_{\mathrm{dc}}$. Figure~2 shows the evolution of the differential resistance with DC current for both the symmetric and Venturi-geometry channels for single layer. In the symmetric channel, the $r_d(I_{\mathrm{dc}})$ curves exhibit the expected symmetry with respect to the direction of the applied current, with a pronounced peak near zero bias that is nearly identical for positive and negative current. In contrast, the Venturi-geometry channel displays a pronounced asymmetry in the differential-resistance curves, with the response depending strongly on the sign of $I_{\mathrm{dc}}$. This asymmetric behavior is consistent with the broken geometrical symmetry of the Venturi structure and indicates a directional dependence of the transport properties. In the bilayer sample, the asymmetry is more pronounced. We attribute this enhancement primarily to the lower electron density in the bilayer system, which increases the nonlinear coefficient in the hydrodynamic Bernoulli response. This is consistent with the theoretical scaling of the nonlinear term with carrier density and indicates that the observed effect is strongly sensitive to the underlying electronic parameters of the system.
In addition, the symmetric channel in the single-layer device exhibits a pronounced negative differential resistance at low current levels, which we attribute to electron heating. Because the system is in the hydrodynamic regime, one may expect, owing to the Gurzhi effect, a decrease in resistance with increasing electron temperature. However, such behavior is not observed in the present samples as the bath temperature is increased.As established in previous studies, the occurrence of the Gurzhi effect depends sensitively on several competing scattering mechanisms, including phonon, impurity, and boundary scattering. Their combined influence does not necessarily result in a monotonic decrease in resistance with increasing bath temperature~\cite{gusev1,gusev2,gusev3}. In many cases, additional modifications of the device geometry or boundary conditions are required for the direct observation of the Gurzhi effect with temperature, for example by introducing random obstacles that enhance hydrodynamic flow and favor the emergence of Gurzhi-type behavior~\cite{levin2}. In the present sample, no clear Gurzhi effect is observed as a function of bath temperature. By contrast, when a dc heating current is applied, the electron temperature can increase independently of the lattice temperature, leading to a reduction in resistance with increasing current. A similar effect was reported in the pioneering work of de Jong \emph{et al.}~\cite{dejong}.

The absence of a clear Gurzhi decrease with bath temperature therefore does not contradict the current-induced decrease of differential resistance. The bath-temperature dependence reflects a competition between electron--electron scattering, phonon momentum relaxation, impurity scattering, boundary scattering, slip length, and the relaxation of different angular harmonics of the distribution function. Under finite dc bias, however, the electron system can be heated out of equilibrium while the lattice remains close to the bath temperature. This can enhance electron--electron scattering without producing the same increase in phonon momentum relaxation as in a bath-temperature sweep. Thus, a current-induced viscous reduction of resistance can appear even when the conventional bath-temperature Gurzhi effect is weak or not clearly resolved. More generally, hydrodynamic transport should be identified from a set of mutually consistent signatures rather than from the Gurzhi effect alone.

In contrast, in the bilayer device we observe neither a Gurzhi effect as a function of temperature nor a pronounced decrease in differential resistance with dc current. It is important to note that the symmetric channel in the bilayer sample was approximately \(20~\mu\mathrm{m}\) long, whereas in the single-layer sample the corresponding channel length was only \(3~\mu\mathrm{m}\). More generally, the longer channel segments studied in our samples have consistently failed to show a clear Gurzhi effect. We attribute this difference to the combined role of boundary conditions, momentum relaxation, and nonequilibrium electron heating. In the hydrodynamic regime, a decrease of resistance requires not only sufficiently strong electron--electron scattering, but also boundary conditions favorable for the development of a Poiseuille-like flow profile.

In the single-layer device, the applied dc current raises the electron temperature and leads to a pronounced decrease in the differential resistance, giving rise to negative differential resistance at low bias. This behavior indicates that current-induced heating drives the system into a regime where viscous effects reduce the resistance. In the bilayer device, however, no comparable decrease is observed. Therefore, the absence of a noticeable reduction of \(r_d\) with increasing dc current in our bilayer sample should not be regarded as a generic property of bilayer systems. Rather, it indicates that, in this particular device, the combined effect of boundary scattering, momentum relaxation, and nonequilibrium heating is not favorable for the emergence of a measurable current-induced viscous reduction of the resistance. Consistently, the longer channel segments in these samples also fail to exhibit the Gurzhi effect.
\begin{figure}[ht]
\includegraphics[width=8cm]{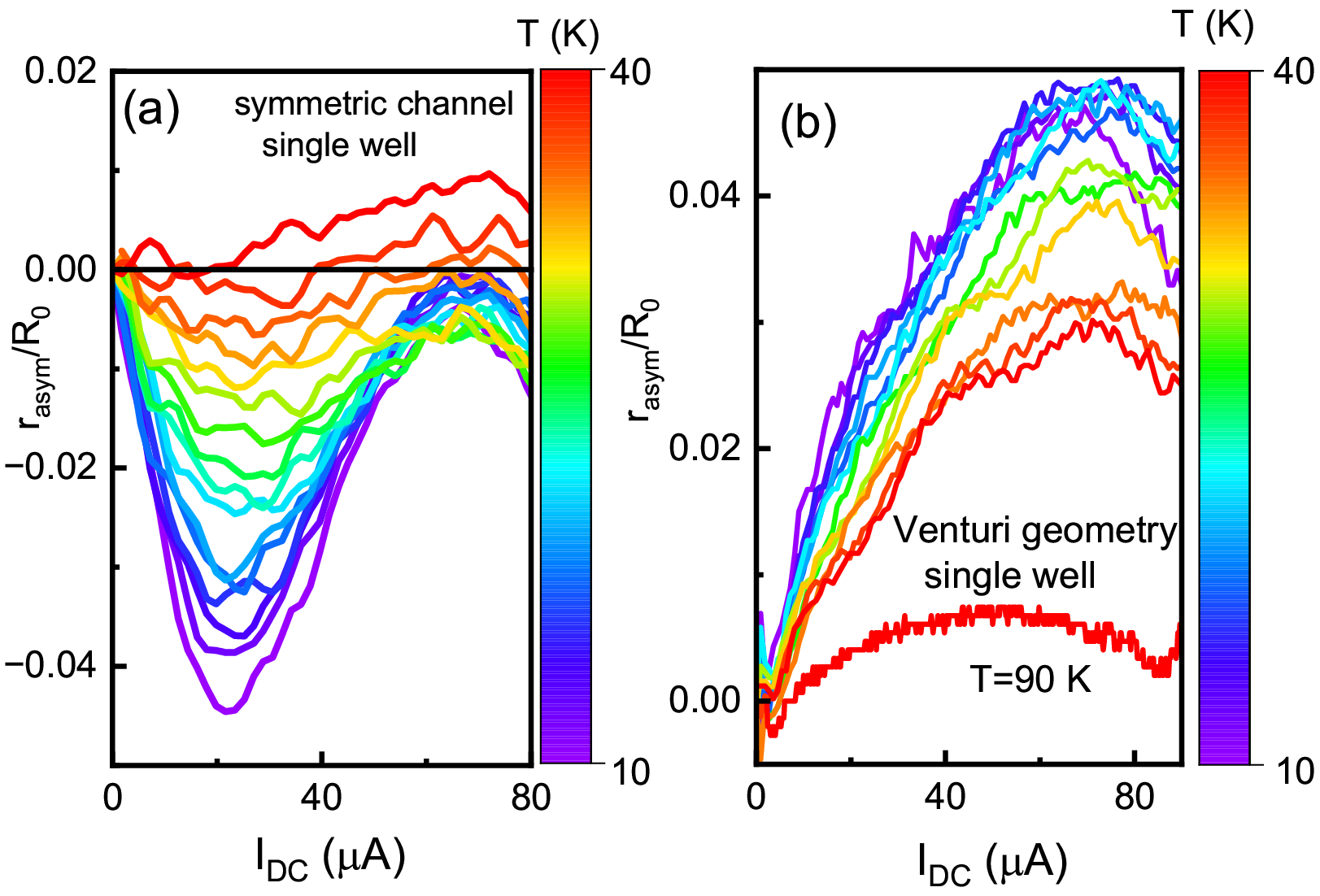}
\caption{\label{sample}(Color online)
Evolution of the normalized left--right differential resistance as a function of applied DC current and temperature for the single-layer device: (a) symmetric channel and (b) Venturi-geometry channel. The red line shows the curve measured at \(T=90~\mathrm{K}\).}
\end{figure}

\begin{figure}[ht]
\includegraphics[width=8cm]{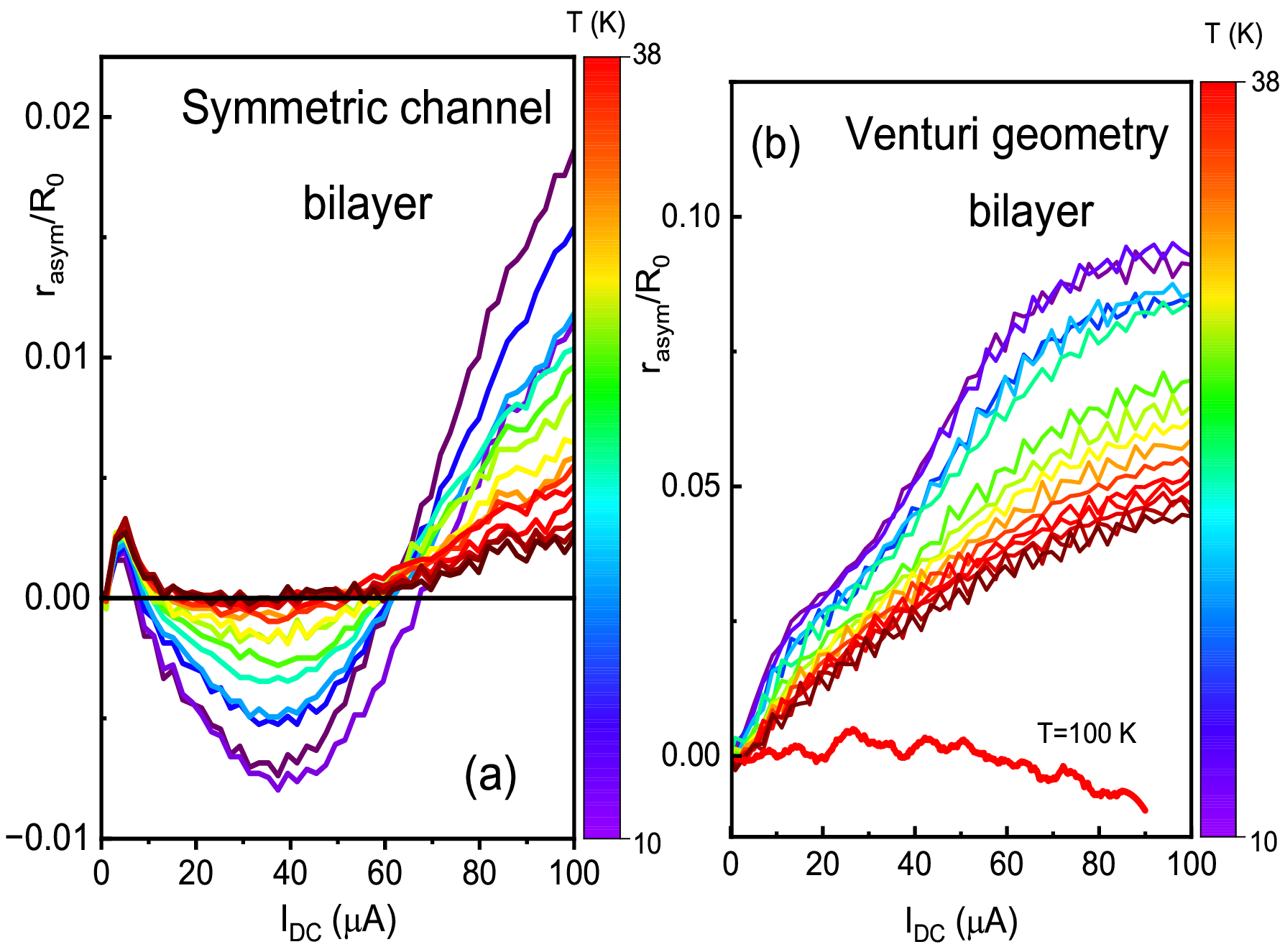}
\caption{\label{sample}(Color online)
Evolution of the normalized left--right differential resistance as a function of applied DC current and temperature for the bilayer device: (a) symmetric channel and (b) Venturi-geometry channel. The red line shows the curve measured at \(T=100~\mathrm{K}\). }
\end{figure}
\begin{figure}[ht]
\includegraphics[width=8cm]{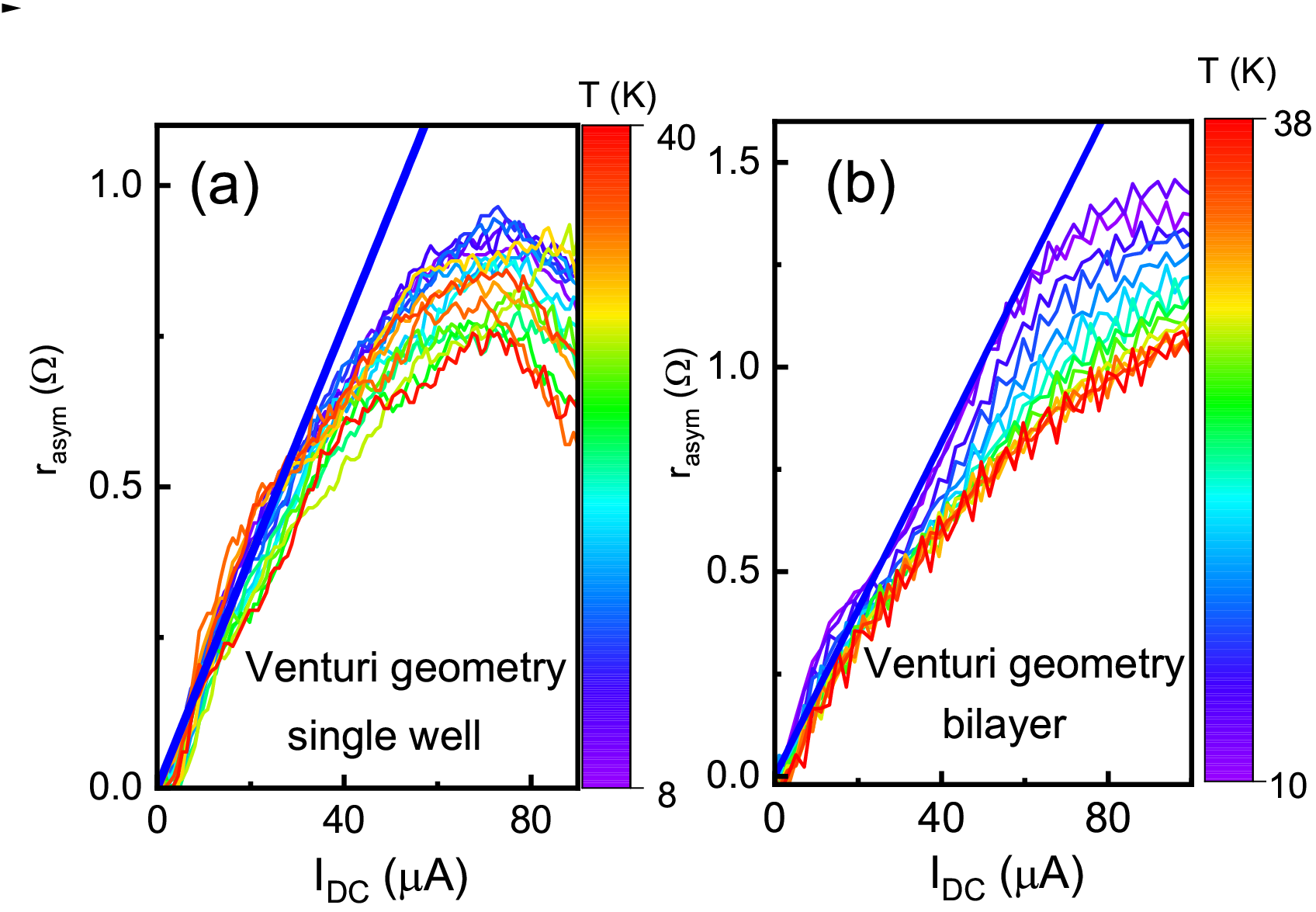}
\caption{\label{sample}(Color online)
 Evolution of the left--right differential resistance as a function of applied dc current and temperature for the single-well sample (a) and the bilayer sample (b). The blue lines show the low-current Bernoulli approximation calculated from Eq.~(10), using \(R_{\rm nl}\simeq 4.3\times10^3~\Omega/{\rm A}\) for the single-well sample and \(R_{\rm nl}\simeq 5.1\times10^3~\Omega/{\rm A}\) for the bilayer sample.}
\end{figure}
Figure~4 shows the left--right differential resistance of the single-layer device as a function of applied DC current and temperature for the symmetric and Venturi-geometry channels. We therefore define the experimentally accessible antisymmetric difference
\begin{equation}
r_{asym}\equiv r(-I)-r(+I),
\label{eq:Delta_r_def}
\end{equation}
For convenience, we present the normalized asymmetry, $r_{\mathrm{asym}}/R_0$, which directly quantifies the strength of the left--right asymmetry relative to the zero-bias resistance $R_0=r(0)$.

The two geometries exhibit qualitatively different behavior. In the symmetric channel [Fig.~4(a)], the signal is relatively weak, shows a nonmonotonic dependence on current, and largely follows the behavior of the differential resistance itself, which suggests that the observed asymmetry is mainly caused by residual sample inhomogeneity. In addition, the residual asymmetry in the symmetric channel is strongly suppressed at elevated temperatures, as expected for a geometrically symmetric structure. The high-temperature curves shown in Figs.~4(b) and 5(b), measured at \(T=90\)~K and \(T=100\)~K, respectively, demonstrate that the Venturi asymmetry is also strongly reduced. At these temperatures, phonon scattering is expected to dominate momentum relaxation, driving the system away from the viscous hydrodynamic regime and toward a more conventional diffusive transport regime. In this regime, a pronounced asymmetric Venturi response is not expected.

We also note that the same applied dc current corresponds to a larger current density in the narrow symmetric channel than in the Venturi geometry. The symmetric channel width is about \(3~\mu{\rm m}\), whereas the characteristic width of the active Venturi region is several times larger. Thus, for the same total current, the current density in the symmetric channel can be larger by a factor of order five, which can enhance current-induced heating and sample-specific nonlinear background features. In contrast, in the Venturi channel [Fig.~4(b)], the left--right differential resistance is strongly enhanced and does not simply reproduce the current dependence of the differential resistance. Instead, it increases nearly monotonically with DC current over a broad temperature range, which is characteristic of hydrodynamic nonlinearity. Only at high current does this trend begin to deviate, most likely due to the onset of electron-heating effects. These results demonstrate that the Venturi-shaped geometry generates a pronounced directional nonlinear response in the single-layer device.
Figure~5 shows the left--right differential resistance of the bilayer device as a function of applied DC current and temperature for both the symmetric and Venturi-geometry channels. The normalized asymmetry, $r_{\mathrm{asym}}(I)/R_0$, exhibits behavior broadly similar to that observed in the single-layer sample, with only minor differences. In particular, the symmetric channel shows a somewhat different resistance profile, whereas the Venturi channel retains an approximately linear current dependence over a wider current range.
We believe that the stronger background in the single-layer sample al lower temperature may indicate larger density inhomogeneity compared with the bilayer sample, possibly related to wafer or growth details.
\begin{figure}[ht]
\includegraphics[width=9cm]{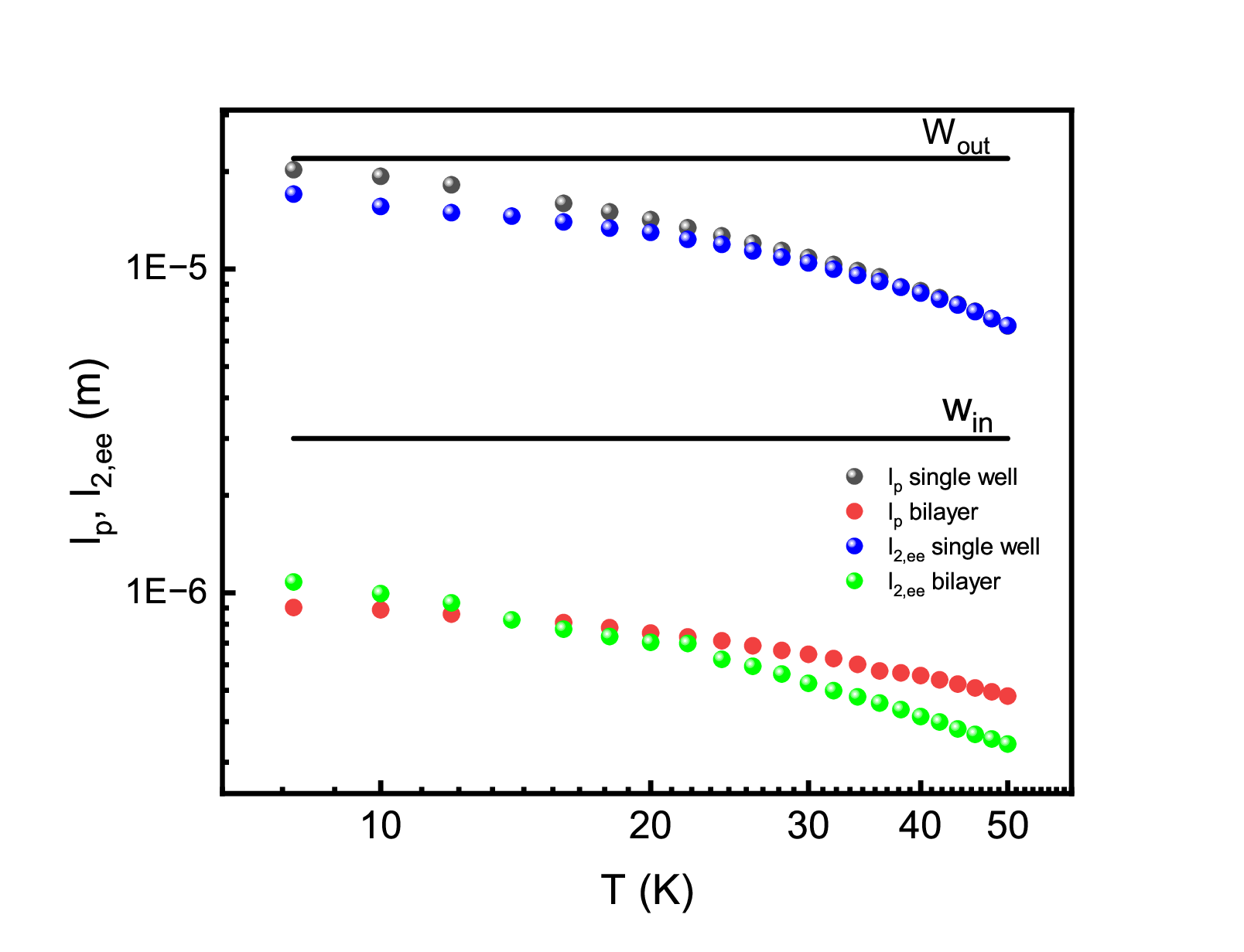}
\caption{\label{bias}(Color online)  Temperature dependence of the extracted scattering lengths \(l_2\) and \(l\) for the single-well and bilayer samples. The horizontal lines indicate the Venturi-device widths, \(W_{\rm in}\) and \(W_{\rm out}\). }
\end{figure}
Hydrodynamic electron flow is expected when the mean free path for electron--electron collisions, \(l_{ee}\), is significantly shorter than the momentum-relaxing mean free path, \(l\), determined by impurity and phonon scattering. To extract these parameters, we performed measurements in the narrow Hall-bar-like channels. This procedure was necessary because the Venturi geometry has a variable width, whereas the viscous magnetotransport model used for comparison with experiment is formulated for a channel of constant width. The measurement results and the extracted parameters are presented in the Supplemental Material~\cite{suppl}. In viscous transport, the relevant quantities include the electron--electron scattering rate $\tau_{2}$ responsible for the relaxation of the second harmonics of the distribution function \cite{alekseev} and the slip length \(l_s\), which controls the boundary condition for electron flow \cite{kiselev}.  Figure~7 shows the temperature dependence of scattering length $l_{2}$ and $l$, where length $l_{2}=v_{F}\tau_{2}$  corresponds to the relaxation of the second harmonic in the distribution function \cite{alekseev}, where $v_{F}$ is the Fermi velocity. Hydrodynamic transport is usually expected when \(l_{ee}<W<l\), where \(W\) is the sample width. At the same time, one should also consider the condition \(l_2<W\), where \(1/\tau_2=v_F/l_2\) characterizes the relaxation of the second angular harmonic of the distribution function. This relaxation is sensitive to disorder and determines the residual relaxation of shear stress as \(T\to 0\). In this language, the viscous hydrodynamic regime corresponds to \(l_2<W<l\). In our samples, this condition is satisfied over a broad temperature range, indicating that the system remains in the viscous transport regime even down to \(T=4.2\)~K~\cite{gusev1,gusev2,levin,levin2,gusev3,levin3}. At the same time, the more restrictive condition based solely on electron--electron scattering is expected to hold only in an intermediate temperature interval, approximately \(25~\mathrm{K}<T<50~\mathrm{K}\). In the Venturi samples, the situation is more complicated because the channel width changes gradually from \(W_{\rm in}\) to \(W_{\rm out}\). In the ideal hydrodynamic limit, the hierarchy \(l_2<W<l\) should hold throughout the entire structure. In practice, however, at \(T=10\)~K the momentum-relaxing mean free path \(l\) is already comparable to \(W_{\rm out}\), and at higher temperatures it becomes significantly shorter. This restricts the range over which the ideal theoretical model discussed below can be applied quantitatively. However, this does not necessarily imply that the nonlinear Venturi response should disappear immediately when \(l<W_{\rm out}\). At elevated temperatures, the relevant hydrodynamic length scale may be set not by the full lithographic width \(W_{\rm out}\), but by a smaller effective region where the current density and velocity gradients are largest, for example near the constriction. Therefore, the effect can be more robust with respect to temperature and may survive even when the momentum-relaxing mean free path becomes shorter than the widest part of the Venturi channel.

\section{Nonlinear Venturi response}

We describe transport in the Venturi-shaped channel within the hydrodynamic Bernoulli framework following Ref.~\cite{hui2}. For a wedge-shaped channel of length $l_r=r_1-r_0$, with widths $h_0=\theta_0 r_0$, $h_1=\theta_0 r_1$, the current--voltage characteristic can be written as
\begin{equation}
V(I)=R_0 I - R_{\mathrm{nl}} I^2,
\label{eq:VI}
\end{equation}
where $R_0$ is the linear-response resistance and $R_{\mathrm{nl}}$ is the leading nonlinear coefficient. 

For the Venturi geometry,
\begin{equation}
R_0=
\frac{1}{\sigma_D}\,
\frac{l_r\,\ln(h_1/h_0)}{h_1-h_0},
\label{eq:R0_sigma}
\end{equation}
and
\begin{equation}
R_{\mathrm{nl}}=
\frac{1}{\sigma_D}\,
\frac{1}{2}
\left(
\frac{1}{h_0^2}-\frac{1}{h_1^2}
\right)
\frac{1}{\rho_e\gamma},
\label{eq:Rnl_sigma}
\end{equation}
where $\sigma_D$ is the Drude conductivity, $\rho_e=ne$ is the two-dimensional charge density, $n$ is the carrier density, and $\gamma$ is the momentum-relaxation rate. Using the Drude expression $\sigma_D=\frac{ne^2}{m^\ast\gamma}$,
we obtain
\begin{equation}
R_0=
\frac{m^\ast\gamma}{ne^2}\,
\frac{l_r\,\ln(h_1/h_0)}{h_1-h_0},
\label{eq:R0_drude}
\end{equation}
and
\begin{equation}
R_{\mathrm{nl}}=
\frac{m^\ast}{2e\rho_e^2}
\left(
\frac{1}{h_0^2}-\frac{1}{h_1^2}
\right)
=
\frac{m^\ast}{2n^2e^3}
\left(
\frac{1}{h_0^2}-\frac{1}{h_1^2}
\right).
\label{eq:Rnl_final1}
\end{equation}
Equation~\eqref{eq:Rnl_final1} shows that the nonlinear coefficient is independent of the momentum-relaxation rate $\gamma$, whereas the linear resistance retains the conventional Drude dependence on $\gamma$. 

In experiment, the relevant quantity is the differential resistance
\(r(I)\equiv dV/dI\). Differentiating Eq.~\eqref{eq:VI} gives
\begin{equation}
r(I)=R_0-2R_{\mathrm{nl}}I,
\label{eq:rI}
\end{equation}
so that
\begin{equation}
r(+I)=R_0-2R_{\mathrm{nl}}I,
\qquad
r(-I)=R_0+2R_{\mathrm{nl}}I.
\label{eq:r_pm}
\end{equation}
We define the experimentally accessible antisymmetric differential resistance as
\begin{equation}
r_{\rm asym}\equiv r(-I)-r(+I),
\label{eq:Delta_r_def}
\end{equation}
which yields
\begin{equation}
\ r_{\rm asym}=4R_{\mathrm{nl}}I.
\label{eq:Delta_r_abs}
\end{equation}
Thus, at low bias the antisymmetric part of the differential resistance is expected to depend linearly on the applied current. Importantly, in the ideal Bernoulli model the nonlinear coefficient \(R_{\mathrm{nl}}\) is determined by the carrier density and the device geometry and is independent of the momentum-relaxation rate \(\gamma\). Therefore, the absolute asymmetry \( r_{\rm asym}\), rather than the normalized quantity \(r_{\rm asym}/R_0\), provides the most direct comparison with the theoretical prediction.

In Figs.~6(a) and 6(b), we plot \(r_{\rm asym}\) as a function of dc current for the Venturi geometry. This allows a direct comparison between experiment and theory without normalizing by the zero-current resistance \(R_0\). In both the single-well and bilayer devices, the data show an approximately linear dependence at low current, as expected from Eq.~\eqref{eq:Delta_r_abs}. Moreover, the absolute asymmetry is only weakly temperature dependent over the temperature range at low current density. This behavior is consistent with the ideal Bernoulli prediction, in which \(R_{\mathrm{nl}}\) is insensitive to the momentum-relaxation rate.  Moreover, as discussed above, the effect is relatively robust with respect to temperature and can remain observable even when \(l<W\).

We now compare the extracted nonlinear coefficient with the theoretical estimate. From the low-current slope of the absolute antisymmetric differential resistance, shown by the blue lines in Figs.~6(a) and 6(b), \(dr_{\rm asym}/dI=4R_{\rm nl}\), we obtain
\(R_{\rm nl}^{\rm exp}\simeq4.3\times10^3~\Omega/{\rm A}\)
for the single-well sample and
\(R_{\rm nl}^{\rm exp}\simeq5.1\times10^3~\Omega/{\rm A}\)
for the bilayer sample.

For comparison with the theory of Ref.~\cite{hui2}, we use \(W_{\rm in}\), \(W_{\rm out}\), and \(l_r\) as the effective parameters of the active Venturi region. The ideal model estimate based on Eq.~\eqref{eq:Rnl_final1} can be substituted by 
\begin{equation}
R_{\mathrm{nl}}=\frac{m^\ast}{2n^2e^3}
\left(
\frac{1}{W_{in}^2}-\frac{1}{W_{out}^2}
\right).
\label{eq:Rnl_final}
\end{equation}
In this case \(R_{\rm nl}\) is very sensitive to the effective width of the narrow Venturi throat. Therefore, small deviations from the lithographic width, caused by current spreading, finite slip, edge depletion, or a nonideal velocity profile near the constriction, can noticeably suppress the nonlinear coefficient. Using geometric  \(W_{\rm out}=22~\mu{\rm m}\) (see Fig.~1), we find that an effective throat width \(W_{\rm in}^{\rm eff}\simeq4.5~\mu{\rm m}\) gives
\(R_{\rm nl}\simeq4.3\times10^3~\Omega/{\rm A}\)
for the single-well sample, while \(W_{\rm in}^{\rm eff}\simeq5.5~\mu{\rm m}\) gives
\(R_{\rm nl}\simeq5.1\times10^3~\Omega/{\rm A}\)
for the bilayer sample. These values are close to the experimental estimates and should be regarded as reasonable agreement with the Bernoulli theory once the effective Venturi geometry is taken into account.

The effective geometry should not be regarded as unique. As shown in Fig.~7, at elevated temperatures the momentum-relaxing mean free path becomes shorter than the full lithographic width \(W_{\rm out}=22~\mu\mathrm{m}\), while the low-current slope of \(\Delta r_{\rm asym}\) remains only weakly temperature dependent. This suggests that the nonlinear response is governed by an effective active region: the throat width is larger than the lithographic \(W_{\rm in}=3~\mu\mathrm{m}\), whereas the effective outer width may be smaller than the full \(W_{\rm out}\). To avoid introducing two independent effective widths, we keep \(W_{\rm out}=22~\mu\mathrm{m}\) fixed and describe the nonideal current focusing through \(W_{\rm in}^{\rm eff}\). This choice gives a reasonable estimate of \(R_{\rm nl}\) and also leads to a consistent estimate of \(R_0\).

The corresponding linear resistance should be discussed separately. In contrast to \(R_{\rm nl}\), which is very sensitive to the effective throat width, \(R_0\) is controlled mainly by the effective zero-field resistivity of the narrow mesoscopic channel. A detailed estimate of \(R_0\), based on the momentum-relaxation and viscous parameters extracted from the magnetoresistance measurements, is presented in ~\cite{suppl}. This analysis gives values close to the measured zero-bias resistances \(R_0\simeq15\text{--}16~\Omega\). We note that this estimate is independent of the Venturi nonlinear-response fit, since it uses the transport parameters extracted from magnetoresistance measurements in the Hall-bar-like channel.

At higher bias, systematic deviations from the low-current linear dependence become visible. This regime also exhibits a stronger temperature dependence, in contrast to the low-current response, which remains only weakly temperature dependent. These deviations are most naturally attributed to electron heating. As the dc current increases, the electron temperature rises, and the ideal low-current expansion is no longer expected to contain only the leading Bernoulli term. Although the momentum-relaxation time does not enter the ideal expression for \(R_{\rm nl}\), heating can modify the viscosity, slip length, boundary conditions, and current distribution in the real device. Equivalently, the apparent nonlinear coefficient becomes current and temperature dependent, \(R_{\rm nl}(I,T_e)\). This can generate higher-order odd-in-current corrections to \(r_{\rm asym}\).

This interpretation is supported by a simple estimate based on the electron thermometry established for the same GaAs material system in Ref.~\cite{levin4}. In the current range where the nonlinear response becomes pronounced, the electron temperature is expected to reach $T_e\sim 30$--$40$ K, while the lattice remains close to the bath temperature. Since the electron-electron shear-stress relaxation rate scales approximately as $1/\tau_{2,ee}\propto T_e^2$, this corresponds to an increase of about one order of magnitude relative to its value at $T=10$ K. By contrast, the phonon momentum-relaxation contribution $1/\tau_{0,ph}=B_{\rm ph}T_{\rm lattice}$ remains essentially fixed by the lattice temperature. Thus, DC bias can selectively enhance the relaxation of the second angular harmonic of the distribution function without producing the same increase in phonon momentum relaxation as a bath-temperature sweep.

It is also useful to discuss this point in terms of the Reynolds numbers introduced in Ref.~\cite{hui2}. In that theory, nonlinear hydrodynamics is characterized by two dimensionless parameters: the viscous Reynolds number, \({\rm Re}_{\nu}=vL/\nu\), and the momentum-relaxation Reynolds number, \({\rm Re}_{\gamma}=v/(L\gamma)\). The analysis of Ref.~\cite{hui2} is developed in the low-Reynolds-number regime, \({\rm Re}_{\nu},{\rm Re}_{\gamma}\ll1\). Here, \(v\) is the characteristic drift velocity, \(L\) is the relevant geometrical length scale, \(\nu\) is the kinematic viscosity, and \(\gamma\) is the momentum-relaxation rate.

Using the parameters extracted from the magnetoresistance measurements shown in Fig.~7, we estimate these quantities for the present devices. With these values, the momentum-relaxation Reynolds number remains small: \({\rm Re}_{\gamma}\simeq1.4\times10^{-2}\) for the single-well sample and \({\rm Re}_{\gamma}\simeq1.8\times10^{-2}\) for the bilayer sample at \(I=20~\mu{\rm A}\). Even at the highest applied currents, \({\rm Re}_{\gamma}\) remains well below unity. Therefore, the high-current deviation from the low-bias linear dependence is unlikely to originate primarily from a breakdown of the low-\({\rm Re}_{\gamma}\) approximation.

The viscous Reynolds number is of order unity at \(I=20~\mu{\rm A}\): \({\rm Re}_{\nu}\simeq0.75\) for the single-well sample and \({\rm Re}_{\nu}\simeq1.05\) for the bilayer sample. Since \({\rm Re}_{\nu}\) grows linearly with current, it can exceed unity at higher bias, particularly in the bilayer case. However, the experimental trend does not indicate that the high-bias deviations are primarily controlled by the Reynolds number. In particular, the bilayer sample, for which \({\rm Re}_{\nu}\) is comparable to or larger than that of the single-well sample, remains closer to the low-current linear dependence over a wider current range. This trend is more naturally explained by different degrees of electron heating than by a breakdown of the low-Reynolds-number approximation.

Finally, it is worth mentioning that an asymmetric geometry can, in principle, generate a directional response in a purely ballistic regime. However, based on the magnetotransport analysis discussed above, we describe the present experiment as being in the hydrodynamic or viscous regime, with a possible ballistic--hydrodynamic crossover at the lowest temperatures, rather than in a purely ballistic regime. Moreover, a ballistic geometrical asymmetry would not by itself explain the specific nonlinear dependence observed in our data. Ballistic transmission through an asymmetric structure may produce a directional conductance difference, but such an effect is not expected, without an additional mechanism, to generate a Bernoulli-type differential response in which \(\Delta r_{\rm asym}\) is proportional to current.

At present, we are not aware of a ballistic theory for the present Venturi geometry that predicts the observed low-bias linear antisymmetric differential resistance, its sign, and its magnitude. By contrast, the hydrodynamic theory of Ref.~\cite{hui2} predicts precisely such a geometry-dependent nonlinear response in a channel with spatially varying width. Our measurements show the expected behavior: a robust low-bias slope in the Venturi geometry, a much weaker and nonmonotonic background in the symmetric channel, and a nonlinear coefficient of the same order of magnitude as the theoretical estimate.
\section{Conclusion}

In conclusion, we have investigated nonlinear transport in high-mobility two-dimensional electron systems patterned into a Venturi-shaped geometry and compared the results with a symmetric control channel. In both single-well and bilayer samples, the Venturi geometry exhibits a pronounced left--right asymmetry of the differential resistance, whereas the corresponding response in the symmetric channel is absent, strongly suppressed, or appears only as a nonmonotonic background. This clear contrast demonstrates the geometry-dependent nature of the observed nonlinear response.

The hydrodynamic character of the transport regime is further supported by magnetoresistance measurements and the extraction of the relevant relaxation parameters. These measurements show that the samples satisfy the viscous-transport conditions in the temperature range where the nonlinear Venturi response is observed. In addition, high-temperature measurements show that the Venturi asymmetry is suppressed when phonon momentum relaxation becomes dominant, consistent with the hydrodynamic origin of the effect.

Importantly, the present results show that nonlinear hydrodynamic effects can be accessed already at very low momentum-relaxation Reynolds number. This distinguishes the Venturi Bernoulli response from many other nonlinear hydrodynamic phenomena, which typically require substantially higher Reynolds numbers. The Venturi geometry therefore provides a favorable route for observing nonlinear electron hydrodynamics in experimentally realistic mesoscopic devices.

Using the effective Venturi geometry together with the transport parameters extracted independently from magnetoresistance measurements, we obtain a consistent quantitative estimate of both the nonlinear coefficient and the zero-bias resistance. In particular, the effective throat width accounts for the measured \(R_{\rm nl}\), while the inclusion of the viscous boundary contribution to the linear resistance gives \(R_0\) close to the experimental values. This agreement shows that the observed response is quantitatively compatible with the hydrodynamic Bernoulli framework once realistic device geometry and viscous boundary effects are taken into account.

Overall, our results identify the Venturi geometry as a sensitive probe of nonlinear hydrodynamic effects in mesoscopic electron systems. The observed response is consistent with a hydrodynamic Bernoulli-type contribution and highlights the importance of realistic geometry and boundary scattering for quantitative modeling. These findings open a route toward further studies of nonlinear electron hydrodynamics in nonuniform channels, including the ballistic--hydrodynamic crossover and the role of finite-slip boundary conditions in shaping nonlinear flow.

\section {Acknoweldgments}
This work is supported by FAPESP (São Paulo Research Foundation) Grants No. 2019/16736-2, No. 2021/12470- 8, No. 2025/22786-3, CNPq (National Council for Scientific and Technological Development). The growth of GaAs quantum wells and preliminary transport measurements were supported by the Ministry of Science and Higher Education of the Russian Federation. C.A.M is supported by CAPES (88887.199336/2025-00)
\section{DATA AVAILABILITY}
The data that support the findings of this article are openly
available \cite{zenodo}.

\section{SUPPLEMENTAl MATERIAL }

\subsection{Hydrodynamic transport regime} 
The hydrodynamic approach to electron behavior in two-dimensional fermionic systems offers a unique perspective that diverges from traditional kinetic theory, revealing fascinating predictions for electron transport, particularly in small-scale samples. A key insight is that, when electron-electron interactions are strong enough, the system can be described by a viscous hydrodynamic framework, allowing for new interpretations of transport phenomena. Recent breakthroughs in materials science, especially in producing exceptionally clean samples, have enabled researchers to systematically explore these hydrodynamic effects across various two-dimensional electronic systems. Hydrodynamic electron flows are anticipated in transport phenomena when the mean free path for electron-electron collisions (denoted as $l_{ee}$) is significantly shorter than the mean free path due to impurity and phonon scattering (represented as $l$). In order to extract these parameters, we performed measurements in the narrow Hall-bar-like channels. This was necessary because the Venturi geometry has a variable width, whereas the viscous magnetotransport model used for comparison with experiment is formulated for a channel of constant width.

\subsubsection{Measurement details}
The single-layer devices were fabricated from high-quality GaAs quantum wells of nominal width 16~nm, with an electron density of approximately $9.0 \times 10^{11}~\mathrm{cm}^{-2}$ at 4.2~K. The corresponding macroscopic wafer exhibited an electron mobility of about $2 \times 10^{6}~\mathrm{cm}^2/\mathrm{V\,s}$. The bilayer devices were fabricated from wide GaAs quantum wells, in which charge redistribution across the well forms a bilayer electron system with two occupied subbands separated by a soft electrostatic barrier. For such bilayer structures, a representative wide-well sample has a width of 46~nm, a total electron density of $6.7 \times 10^{11}~\mathrm{cm}^{-2}$, and a mobility of about $2 \times 10^{6}~\mathrm{cm}^2/\mathrm{V\,s}$ at 4.2~K.  The sample consists of several different segments, including a Venturi-shaped region and a narrow Hall-bar-like channel. The measurements were performed using the device shown in Fig. 1 of the main text, specifically a short Hall-bar segment with a length of $8 \mu\mathrm{m}$ and a width of $3 \mu\mathrm{m}$. We also integrated eight voltage probes into this setup. Ohmic contacts to the two-dimensional electron system were created by annealing Ti/Ni/Au layers deposited on the GaAs surface. For our measurements, we used a VTI cryostat combined with a standard lock-in detection technique to measure longitudinal resistance. To prevent overheating, we applied an alternating current (AC) in the range of $0.1-1 \mu A$, a level considered sufficiently low for these tests. 

Figure 1 illustrates the variation in resistivity ($\rho = W/L \times R$) of the single well with respect to magnetic field strength at different temperatures. A notable characteristic in these samples is the pronounced negative magnetoresistivity, $\rho(B) - \rho(0) < 0$, which follows a Lorentzian profile. As temperature increases, this negative magnetoresistivity diminishes in magnitude and broadens. Additionally, the resistivity at zero magnetic field rises with temperature.
\begin{figure}[ht]
\includegraphics[width=8cm]{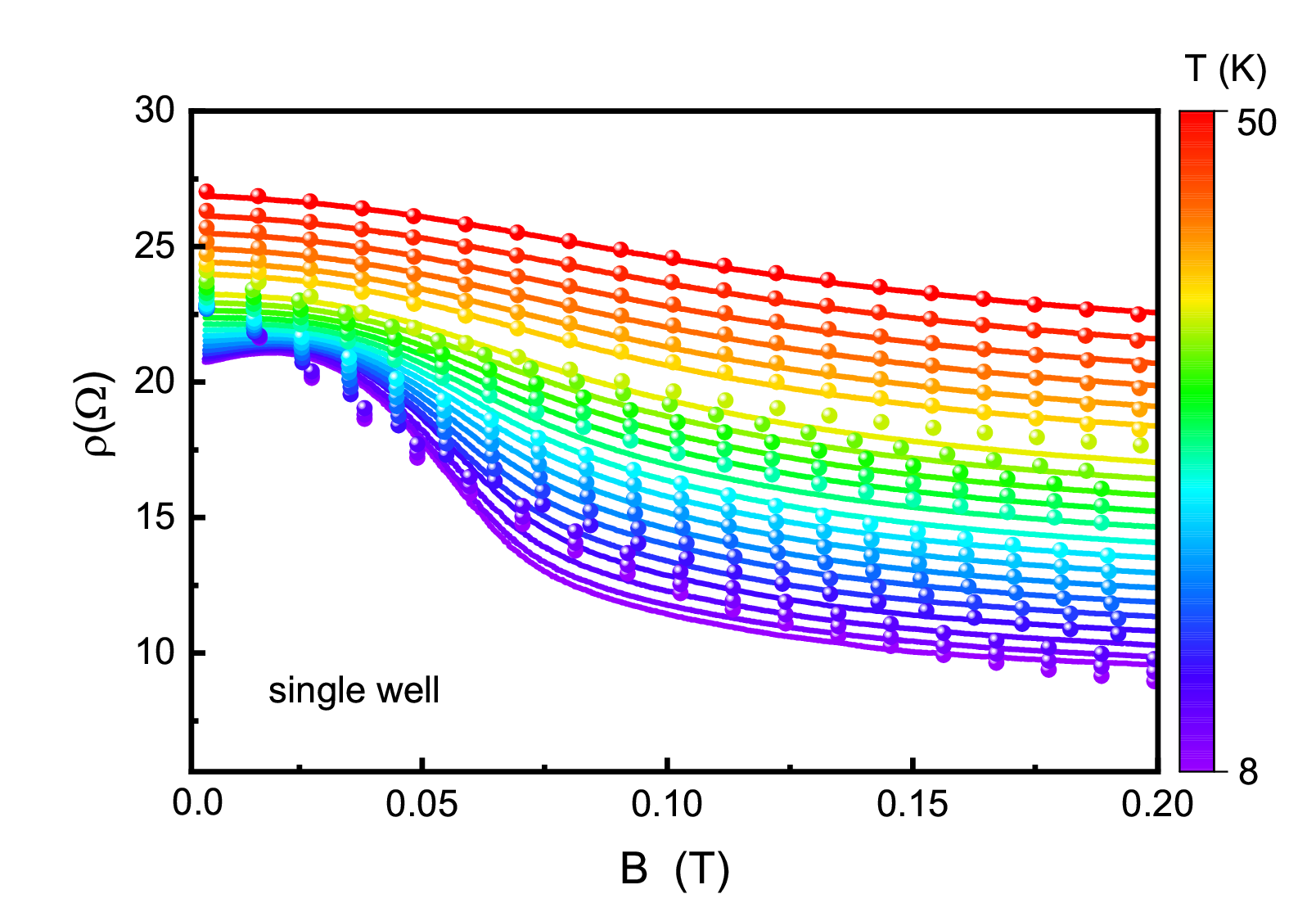}
\caption{\label{fig4}(Color online) Temperature-dependent magnetoresistivity  experiment (solid lines) and  theoretical curves ( circles) for single well calculated from Eq. (1) for different temperatures.}
\end{figure}
\begin{figure}[ht]
\includegraphics[width=8cm]{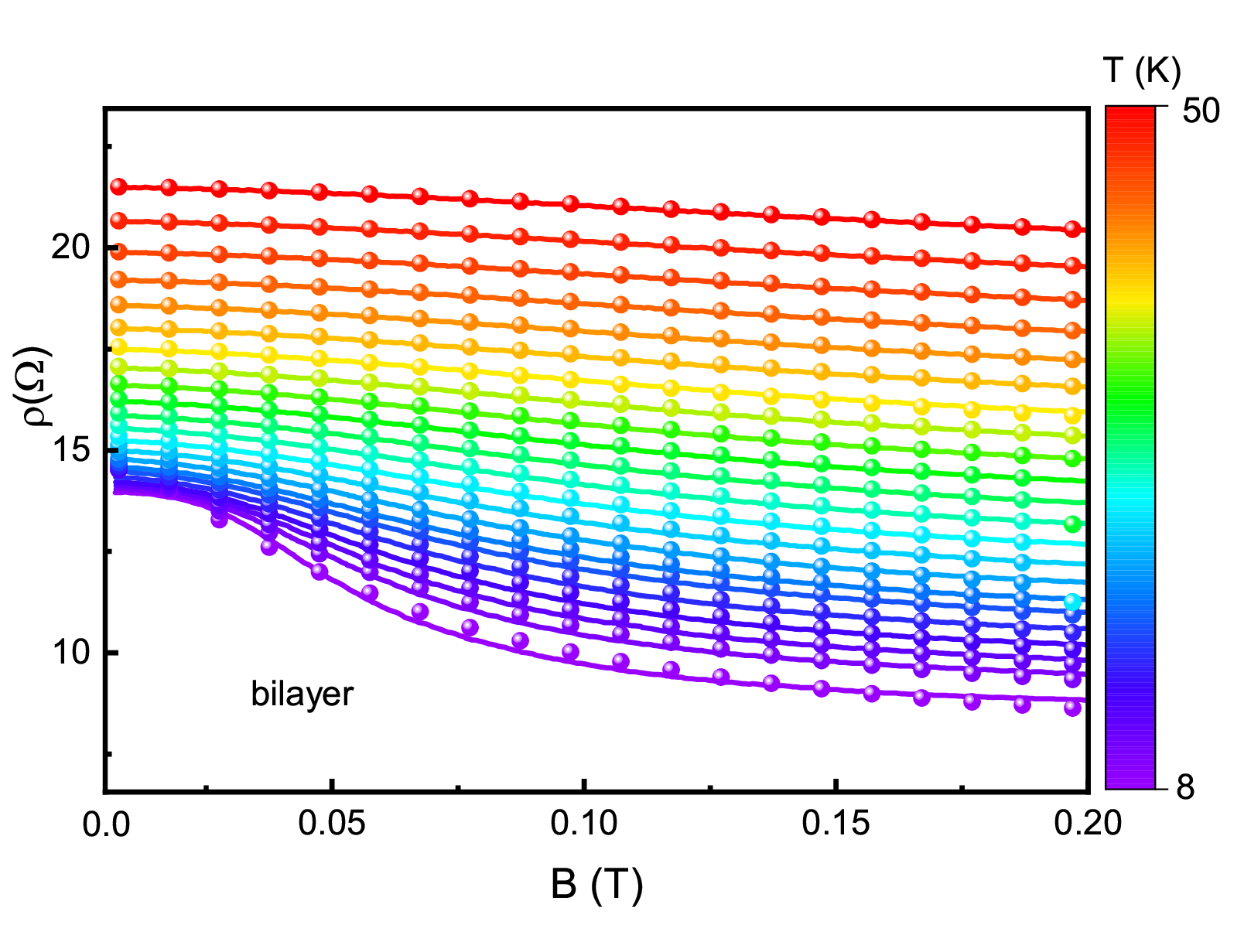}
\caption{\label{fig4}(Color online) Temperature-dependent magnetoresistivity  experiment (solid lines) and  theoretical curves ( circles) for bilayer calculated from Eq. (1) for different temperatures.}
\end{figure}
\subsection{Theory and discussions}
To qualitatively compare with the experimental data from samples without obstacles, we apply a model from previous research, initially designed to describe Poiseuille flow under the influence of a magnetic field \cite{alekseev1}. 
\begin{figure}[ht]
\includegraphics[width=9cm]{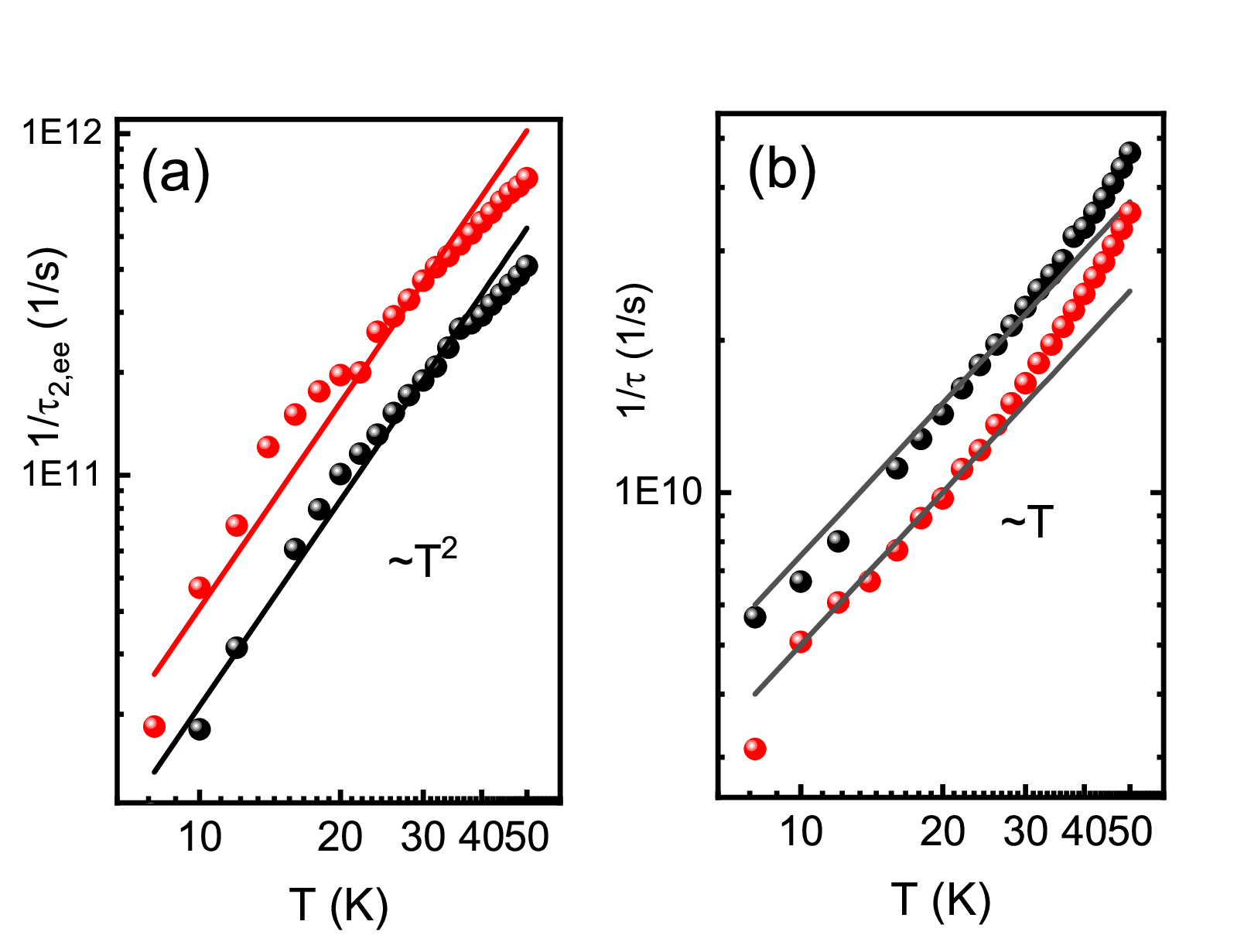}
\caption{\label{fig5}(Color online) (a) $1/\tau_{2,ee}$ (black circles) as a function of the temperature for single well (black circles) and bilayer( red circles). (b) The relaxation rate $1/\tau$ as a function of the temperature for single well (black circles) and bilayer( red circles). }
\end{figure}

For a more complete formulation in magnetic field, the theory incorporates a viscosity tensor, which is dependent on the magnetic field, to determine the resistivity tensor:

\begin{equation}
\rho(B)=\left(\frac{m}{e^{2}n\tau}\right)\frac{1}{1-\tanh(\xi)/\xi}
\end{equation}
In this context, the dimensionless Gurzhi parameter is defined as $\xi = \xi_{0} \sqrt{1 + (2l_{2}/r_c)^2}$, where $\xi_{0} = W^{*}/l_{G}$, with $l_{G} = \sqrt{l_{2}l}$ representing the Gurzhi length. Here, $l_{2} = v_{F} \tau_{2}$, $l = v_{F} \tau$, and $r_{c} = v_{F}/\omega_{c}$ is the cyclotron radius. The cyclotron frequency is $\omega_{c} = eB/mc$. The relaxation rate $\tau_{2}$ corresponds to the shear stress relaxation time arising from electron-electron scattering. The subscript "2" signifies that the viscosity coefficient is governed by the relaxation of the second harmonic in the distribution function \cite{alekseev1}. The shear viscosity relaxation rate is given by 
\begin{equation}
1/\tau_{2}(T) = 1/\tau_{2,ee}(T) + 1/\tau_{2,imp} 
\end{equation}
while the momentum relaxation rate is expressed as
\begin{equation}
1/\tau(T) = 1/\tau_{0,ph}(T) + 1/\tau_{0,imp}
\end{equation}
In this expression, $1/\tau_{0,ph} = B_{ph}T$ corresponds to phonon scattering, and $1/\tau_{0,imp}$ represents scattering due to static disorder, distinct from the relaxation time for the second harmonic \cite{alekseev1}.

We then fit the magnetoresistance curves using Eq. (1). At each temperature, the full measured $\rho(B)$ curve, including the (B=0) point, is fitted directly as a function of magnetic field. The fitting procedure employs three adjustable parameters: the momentum relaxation time $\tau(T)$, the second-harmonic relaxation time $\tau_2(T)$, and the effective hydrodynamic channel width $W^*$.

Figures 1 and 2 show a direct comparison between the experimental magnetoresistance curves and the calculated curves obtained from Eq. (1). For the single-well sample, the agreement with the hydrodynamic model is generally good. However, at the lowest temperatures, the experimental curves show small deviations from the calculated line shape. These deviations are likely related to residual ballistic contributions, which become more important at low temperature as the relevant mean free paths increase. A fully quantitative description in this regime would require treating ballistic and hydrodynamic contributions on the same footing. In the present analysis, however, we use the hydrodynamic expression as a simplified fitting model.

For the bilayer sample, the agreement between experiment and theory is good over the full temperature interval studied. This indicates that the bilayer device provides more favorable conditions for the hydrodynamic description in the present magnetotransport analysis.  Figure 3 illustrates the temperature-dependent behavior of the corresponding relaxation rates. To facilitate comparison with theoretical predictions, we used the parameters \( \frac{1}{\tau_{2,imp}} \), \( \frac{1}{\tau_{0,imp}} \), \( A_{ee} \), and \( B_{ph} \), as listed in Table 1. Employing Eq. (3), the e-e relaxation rate is expressed as
\begin{equation}
\frac{\hbar}{\tau_{2,ee}} = A_{ee} \frac{(kT)^2}{E_F}
\end{equation}
\begin{table}[ht]
\caption{\label{tab1} Fitting parameters of the electron system. Parameters are defined in the text.}
\begin{ruledtabular}
\begin{tabular}{lcccccc}
& Device& $1/\tau_{2,imp}$&$1/\tau_{0,imp}$ & $A_{ee}$ & $B_{ph}$ &$W^{*}$   \\
& &  $(10^{11} 1/s)$ & $(10^{10} 1/s)$ &   & ($10^{9} 1/sK$) & $\mu m$  \\
\hline
&Single well & $4.45$  & $1.45$ & $0.6$ &  $0.75$ & 4.5\\
&bilayer & $3.15$  & $1.8$ & $0.9$ &  $0.5$ & 8\\
\end{tabular}
\end{ruledtabular}
\end{table}
It can be observed that all relaxation rates  converge onto universal curves: \( \frac{1}{\tau_{2,ee}} \sim T^2 \) and \( \frac{1}{\tau} \sim T \). The deviation of \(1/\tau\) at high temperatures may be attributed to the onset of optical phonon emission. One can see also that the effective width is larger than the geometric width. We attribute this discrepancy to the finite slipping length caused by specific scattering at the boundaries. A theoretical model  \cite{gromov} proposes that \( W^{*2} = W(W + 6l_s) \), which indeed predicts a larger effective width for samples with a finite slip length. By comparing with this model, we estimate \( l_s \sim 0.5 \, \mu \mathrm{m} \) for single layer and \( l_s \sim 2 - 4 \, \mu \mathrm{m} \) for bilayer.  For diffusive boundary scattering, the velocity distribution profile in the channel is parabolic, corresponding to Poiseuille flow in a liquid. The slip length is the distance where the extrapolated velocity vanishes \cite{kiselev}. A finite slip length modifies the velocity distribution, shaping it as a "cut parabola $(W+2l_s)$" \cite{kiselev}.

Based on these calculations, we can discuss the conditions for hydrodynamic effects in narrow channel samples. The hydrodynamic description is applicable under conditions where $l_{2,ee} < W < l$, with  $W$ the width of the sample. In this scenario, this condition  indicating that we remain within the hydrodynamic regime  at $T > 25\,\text{K}$.  Additionally, one can expect that in mesoscopic samples, the scattering rate $1/\tau$ is typically lower than in macroscopic samples due to boundary scattering and geometric factors.
 It can be seen that the hydrodynamic condition $l_{2}< W < l$ is satisfied across the entire temperature range used in the experiment. 

In the Venturi samples, the situation is more complicated because the channel width changes gradually from \(W_{\rm in}\) to \(W_{\rm out}\). In the ideal hydrodynamic limit, the hierarchy \(l_{2,ee}<W<l\) should hold throughout the entire structure. In practice, however, at \(T=10\) K the momentum-relaxing mean free path \(l\) is already comparable to \(W_{\rm out}\), and at higher temperatures it becomes significantly shorter. However, this does not necessarily imply that the nonlinear Venturi response should disappear immediately when \(l<W_{\rm out}\). At elevated temperatures, the relevant hydrodynamic length scale may be set not by the full lithographic width \(W_{\rm out}\), but by a smaller effective region where the current density and velocity gradients are largest, for example near the constriction. Therefore, the effect can be more robust with respect to temperature and may survive even when the momentum-relaxing mean free path becomes shorter than the widest part of the Venturi channel.
\subsubsection{Estimate of the linear resistance in the effective Venturi geometry}

The linear resistance of the Venturi region should include not only the ordinary momentum-relaxation contribution, but also the viscous boundary contribution. In the ideal expression of Ref.~\cite{hui2}, the linear resistance has a Drude-like form. However, in a narrow mesoscopic channel the zero-field resistivity contains both the usual momentum-relaxation term and a viscous boundary term. Following the viscous magnetotransport analysis used above,  the effective local resistivity is given by Eq.~1. 
 In the Venturi geometry the width varies from \(W_{\rm in}\) to \(W_{\rm out}\), so the Drude and viscous contributions to the linear resistance can be estimated by integrating along the active region. 
 
For the linear resistance of the Venturi region, it is useful to include both the ordinary momentum-relaxation contribution and the viscous boundary contribution. We estimate the local resistance of a channel segment of width \(W(x)\) as
\begin{equation}
dR=\rho_{\rm eff}(W)\frac{dx}{W(x)} ,
\end{equation}
where
\begin{equation}
\rho_{\rm eff}(W)=
\frac{m^\ast}{ne^2}
\left(
\gamma+\frac{1}{\tau^\ast(W)}
\right).
\end{equation}
For a no-slip Poiseuille profile, the viscous boundary relaxation rate may be estimated as
\begin{equation}
\frac{1}{\tau^\ast(W)}=\frac{12\nu}{W^2},
\end{equation}
where \(\nu\) is the kinematic viscosity. Approximating the Venturi region by a linearly varying width,
\begin{equation}
W(x)=W_{\rm in}+\frac{W_{\rm out}-W_{\rm in}}{l_r}x,
\qquad 0<x<l_r,
\end{equation}
we obtain
\begin{widetext}
    \begin{equation}
R_0=
\frac{m^\ast}{ne^2}
\left[
\gamma
\frac{l_r\ln(W_{\rm out}/W_{\rm in})}
{W_{\rm out}-W_{\rm in}}
+
12\nu
\frac{l_r}{W_{\rm out}-W_{\rm in}}
\frac{1}{2}
\left(
\frac{1}{W_{\rm in}^2}
-
\frac{1}{W_{\rm out}^2}
\right)
\right].
\end{equation}
\end{widetext}
\section{Additional contact configurations}
\begin{figure}[ht]
\includegraphics[width=8cm]{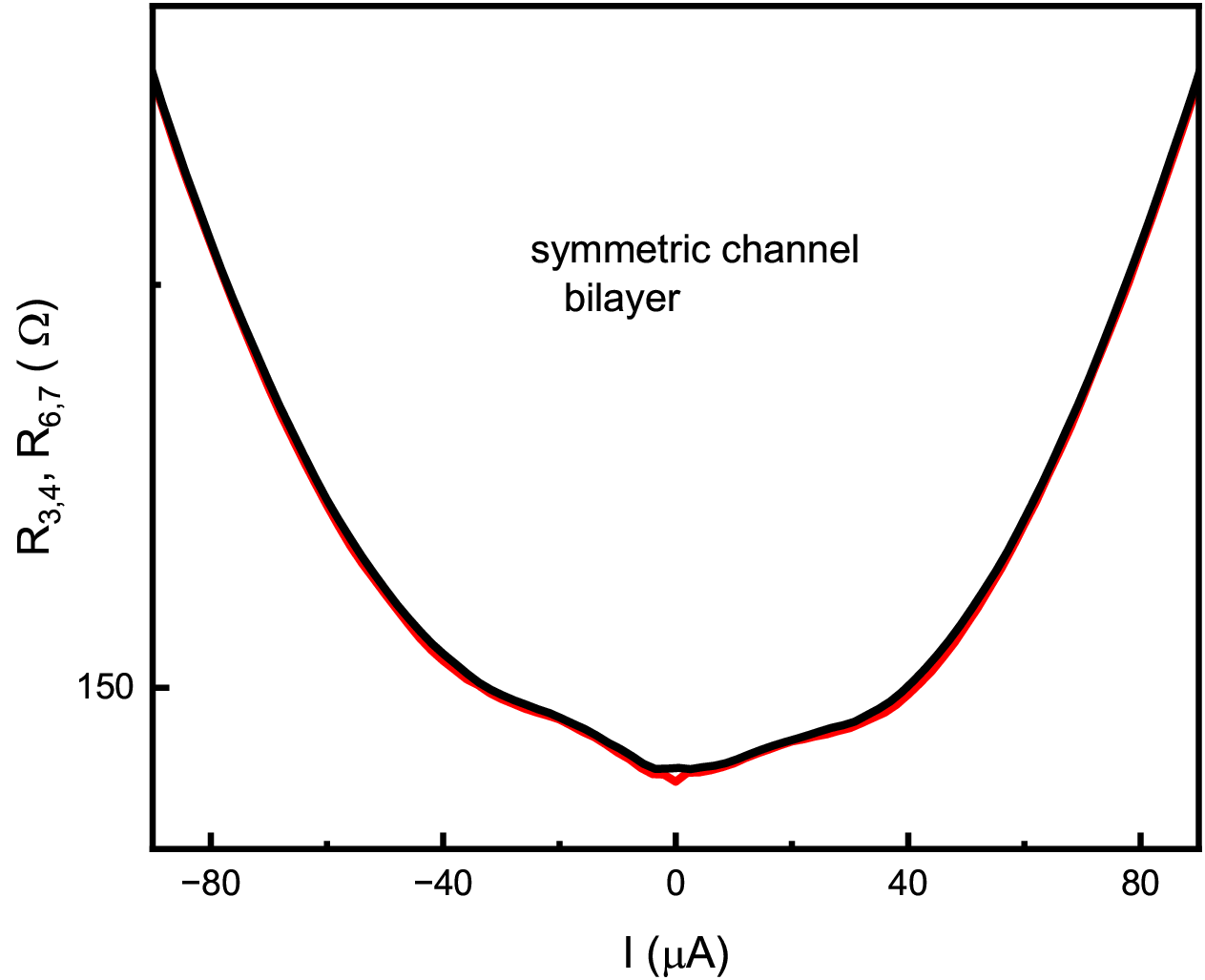}
\caption{\label{sample}(Color online)
 Comparison of differential-resistance measurements obtained using different voltage-contact configurations in the bilayer device at \(T=10~\mathrm{K}\). }
\end{figure}
\begin{figure}[ht]
\includegraphics[width=8cm]{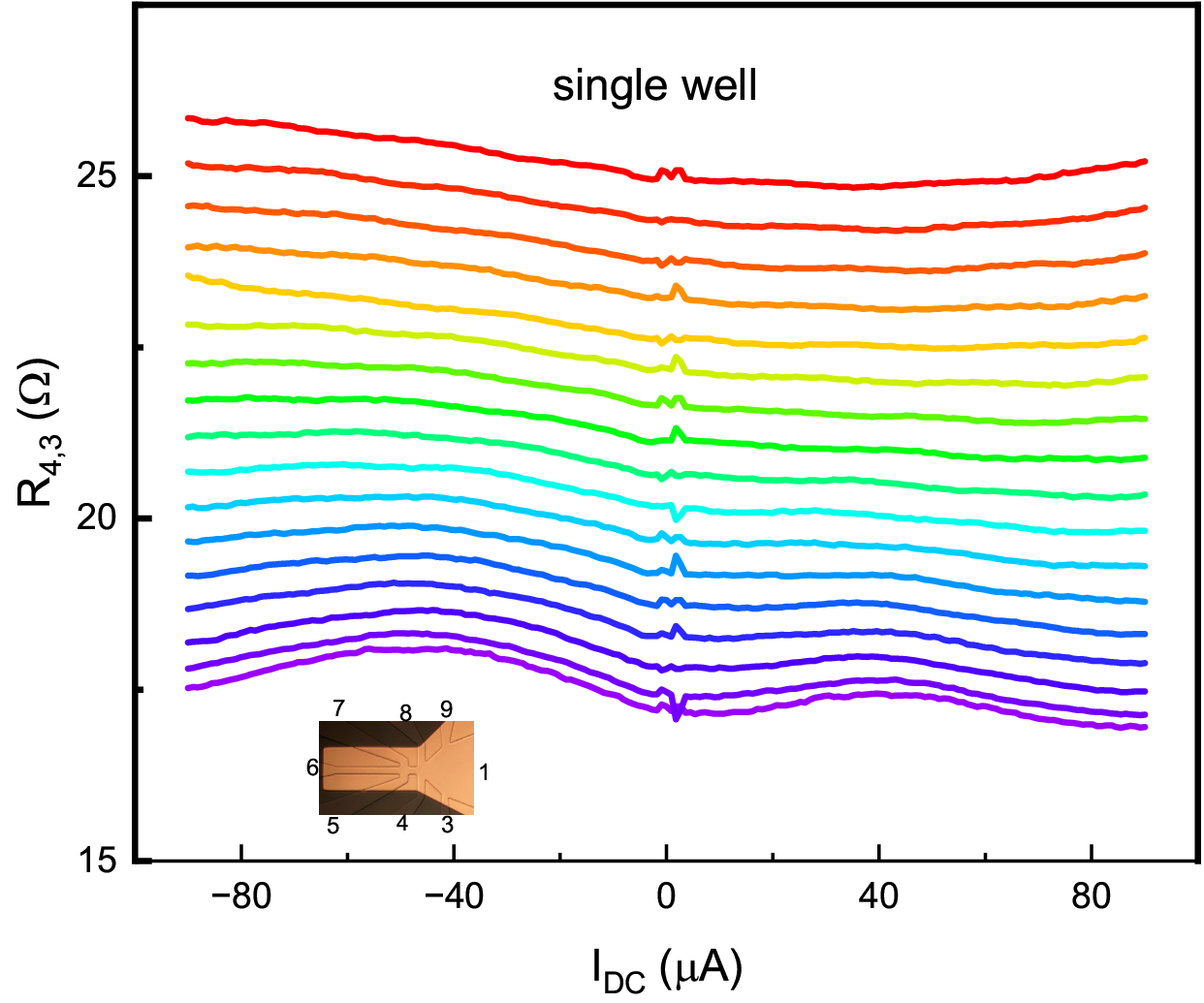}
\caption{\label{sample}(Color online)
 Evolution of the differential resistance as a function of applied DC current and temperature for the Venturi-geometry channel, Single layer }
\end{figure}
Using the viscosity extracted from the magnetoresistance analysis, \(\nu=v_Fl_2/4\), we obtain \(\nu\simeq0.092~{\rm m^2/s}\) for the single-well sample and \(\nu\simeq0.089~{\rm m^2/s}\) for the bilayer sample at \(T=10\)~K. Using \(W_{\rm out}=22~\mu{\rm m}\), \(l_r=15~\mu{\rm m}\), and the effective throat widths inferred from the nonlinear response, \(W_{\rm in}^{\rm eff}\simeq4.28~\mu{\rm m}\) for the single-well sample and \(W_{\rm in}^{\rm eff}\simeq5.51~\mu{\rm m}\) for the bilayer sample, we obtain $R_0\simeq14.4~\Omega$. For the bilayer sample, we obtain $R_0\simeq15.2~\Omega.$

These estimates are close to the measured zero-bias resistances, \(R_0\simeq16~\Omega\) for the single-well sample and \(R_0\simeq15~\Omega\) for the bilayer sample. Thus, once the viscous boundary contribution is included, the effective Venturi geometry gives a consistent estimate of the linear resistance. 

\subsection{Additional contact configurations}

\begin{figure}[ht]
\includegraphics[width=8cm]{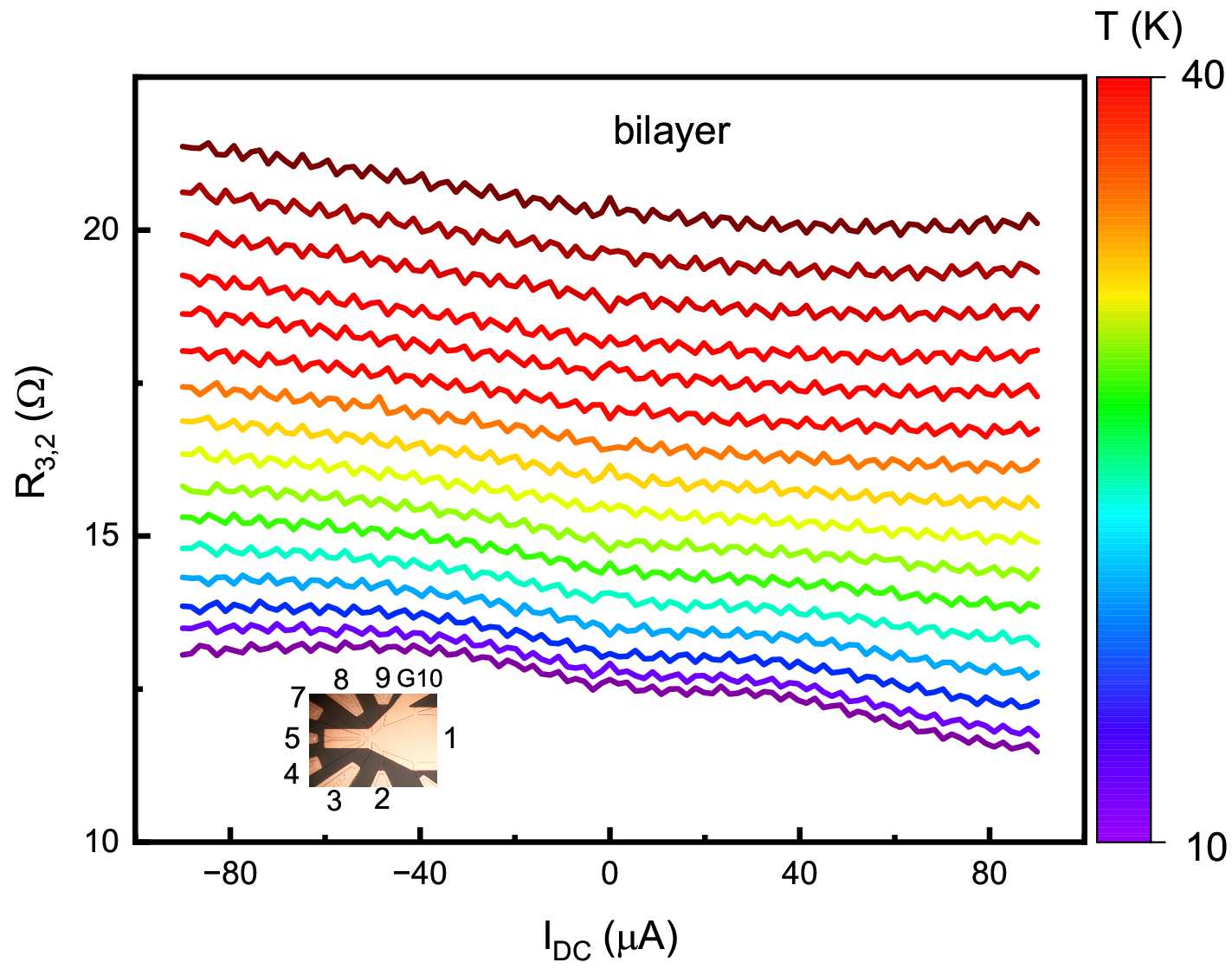}
\caption{\label{sample}(Color online)
 Evolution of the differential resistance as a function of applied DC current and temperature for the Venturi-geometry channel, bilayer}
\end{figure}

To verify that the observed nonlinear response is not specific to a particular voltage-probe pair, we performed additional measurements using different contact configurations. Within experimental accuracy, these measurements give results consistent with those presented in the main text. Figure 4 shows the differential differential resistance as a function of applied dc current for different voltage-contact configurations in the bilayer device at \(T=10~\mathrm{K}\). The similar behavior confirms that the observed response is not specific to a particular contact pair. 

Figures~4 and 5 show the differential resistance as a function of applied dc current at different temperatures for the single-well and bilayer samples, respectively, measured using additional voltage contacts. The overall behavior is similar to that shown in Figs.~2 and 3 of the main text. In particular, the Venturi configuration exhibits the same characteristic left--right asymmetry, while the qualitative temperature and current dependencies remain unchanged.

These results confirm that the observed effect is not caused by a particular choice of voltage probes or by an accidental contact asymmetry.

\end{document}